\documentclass[final,5p,times,twocolumn]{elsarticle}

\usepackage{upgreek}
\usepackage{units}
\usepackage{url}
\usepackage{amsmath}

\usepackage{amssymb}

\usepackage{lineno}

\graphicspath{{../figures_cl/}} 

\journal{Nucl. Instr. Meth. Phys. Res. A}

\newcommand{\hiscore}{HiSCORE}

\begin{document}

\begin{frontmatter}



\title{Event reconstruction techniques for the wide-angle air Cherenkov detector \hiscore{}}


\author{Daniel Hampf}
\ead{daniel.hampf@desy.de}

\author{Martin Tluczykont}
\author{Dieter Horns}

\address{Institute for Experimental Physics, 
University of Hamburg, Luruper Chaussee 149, D-22761 Hamburg, Germany}

\begin{abstract}
Wide-angle, non-imaging air Cherenkov detectors provide a way to observe cosmic gamma-rays which is complementary to observations by imaging Cherenkov telescopes. Their particular strength lies in the multi-TeV to ultra high energy range (E$_\gamma > \unit[30]{TeV}$), where large effective areas, yet small light sensitive areas per detector station are needed. To exploit this potential to full extent, a large station spacing is required to achieve a large effective area at a reasonable effort.

In such a detector, the low number of signals per event, the absence of imaging information, and the poor signal to noise ratio of Cherenkov light to night sky brightness pose considerable challenges for the event reconstruction, especially the gamma hadron separation. The event reconstruction presented in this paper has been developed for the wide-angle detector  \hiscore{}, but the concepts may be applied more generically. It is tested on simulated data in the 10 TeV to 5 PeV energy range using the air shower simulation CORSIKA and the \hiscore{} detector simulation sim\_score. For the tests, a regular grid of $22 \times 22$ detector stations with a spacing of 150 m is assumed, covering an area of 10 km$^2$.

The angular resolution of individual events is found to be about $0.3^\circ$ near the energy threshold, improving to below $0.1^\circ$ at higher energies. The relative energy resolution is 20\% at the threshold and improves to 10\% at higher energies. Several parameters for gamma hadron separation are described. With a combination of these parameters, 80\% to 90\% of the hadronic background can be suppressed, while about 60\% of the gamma-ray events are retained.
The point source sensitivity to gamma-ray sources is estimated, using conservative assumptions, to be about $\unit[8 \times 10^{-13}]{erg\, s^{-1}\, cm^{-2}}$ at $\unit[100]{TeV}$ gamma-ray energy for a 10 km$^2$ array. With more optimistic assumptions, and a 100 km$^2$ array, a sensitivity of about $\unit[1 \times 10^{-13}]{erg\, s^{-1}\, cm^{-2}}$ can be achieved (at $\unit[100]{TeV}$). 

Even in the former case the detector is sensitive enough to measure the continuation of currently known gamma-ray spectra into the ultra high energy domain. Due to its large field of view of $\unit[0.6]{sr}$ it also offers a great potential for the discovery of new gamma-ray sources at the so far largely unexplored energies of 100 TeV and above.

\end{abstract}


\end{frontmatter}


\section{Introduction}

Ground-based gamma-ray astronomy has become a successful and well-established branch of modern astronomy and offers unique possibilities to study the extreme conditions in the non-thermal universe. More than hundred sources of very high energy (VHE,
$\unit[30]{GeV}$ to $\unit[30]{TeV}$) gamma-ray emission are currently known, and detailed spectral, morphological and temporal studies provide a wealth of information on the underlying acceleration and radiation mechanisms of these sources (see e.g.\ \cite{2008RPPh...71i6901A}, \cite{2009NJPh...11k5008K} or \cite{2008RvMA...20..167H} for reviews).

However, at even higher energies, in the ultra high energy band (UHE, $E_\gamma
> \unit[30]{TeV}$), not many sources could be discovered so far. The main
challenge at these energies is the very low photon flux emitted even by strong
gamma-ray sources.  Current instruments for gamma-ray observations, like the
3$^{\mathrm{rd}}$ generation Cherenkov telescope arrays
\cite{2009NJPh...11e5005H}, are usually optimised for a good sensitivity in the
VHE band. While their effective areas of   around $\unit[10^5]{m^2}$
\cite{2006A&A...457..899A} are well suited for this energy band, they are too
small for sensitive observations in the UHE band: For example, in 23 hours of
observations of the Crab Nebula with H.E.S.S., only four gamma-ray events with
energies above $\unit[30]{TeV}$ could be found \cite{2006A&A...457..899A}.

On the other hand, observations at ultra high energies can help to answer
important open questions in high energy astronomy. One of the most important
scientific goals of UHE gamma-ray astronomy is the search for the sources of
charged cosmic rays at PeV energies (PeVatrons) \cite{2007ApJ...665L.131G}. It
is believed that cosmic rays at least up to the so-called knee of their energy
spectrum at around $\unit[4]{PeV}$ are of Galactic origin
\cite{2009PrPNP..63..293B}, but so far no conclusive evidence could be found
for sources or source classes capable of accelerating particles to that energy.
A clear evidence could be a gamma-ray signal above $\unit[100]{TeV}$
\cite{2007ApJ...665L.131G}. However, no such signal could be observed so far,
possibly due to the lack of sensitive observations\footnote{In fact, also a high-energy neutrino signal could be an unambiguous indication for cosmic ray acceleration, but so far none has been found either \cite{2012PrPNP..67..651K}.}. The gamma-ray spectrum at ultra high energies can also be used to distinguish effectively between
leptonic and hadronic cosmic ray accelerators -- a notoriously difficult task
with VHE band observations alone -- since the cross section for leptonic
gamma-ray production decreases strongly in the UHE regime (Klein-Nishina
regime). Further motivations for UHE gamma-ray astronomy can be found in
\cite{Tluczykont20111935, 2008NIMPA.588...48R}.

 
To overcome the challenge of small event numbers in UHE observations, very
large effective areas (in the order of many km$^2$, ideally $\unit[100]{km^2}$
or more) and long exposure times per source are required. 
A powerful concept that can achieve these requirements are arrays of non-imaging, wide-angle air Cherenkov
detectors. In these detectors each station samples a part of the Cherenkov
light front without any image information. Typically, the detector stations
contain only one or a few large photomultiplier tubes (PMTs) directed towards
zenith with a light concentrator on top to increase the light sensitive area
and to restrict the influence of stray light from large zenith angles. The
modest light collection areas of $\mathcal{O}(\unit[1]{m^2})$ (compared to
$\mathcal{O}(\unit[100]{m^2})$ in modern imaging Cherenkov telescopes) are
sufficient since at ultra high energies each incident gamma photon produces a
very large air shower and thereby a relatively strong Cherenkov light pulse. The
inexpensive design of the individual stations and large inter-station distances
make the realisation of large effective areas feasible. Additionally,
individual sources receive on average a higher exposure than from Cherenkov
telescopes, since a large part of the sky is monitored simultaneously (up to 1
sr).  

Experiments of this kind have first been developed for studies of charged
cosmic rays, e.g.\ at Yakutsk \cite{1973ICRC....4.2389D} or at the Buckland
park observatory in Australia \cite{1977ICRC....8..239K}. More recent detectors
employing this method are the BLANKA \cite{2001APh....15...49F} and the TUNKA
\cite{2010arXiv1003.0089B} experiments, the latter still being operational,
again with the focus on charged cosmic rays. The only non-imaging Cherenkov
detector array specifically designed for gamma-ray astronomy has been the
AIROBICC detector built by the HEGRA collaboration \cite{1995APh.....3..321K}.
With an instrumented area of only $\unit[3 \times 10^4]{m^2}$ this array was
however too small to detect gamma-ray sources above its energy threshold of
about $\unit[15]{TeV}$.

Currently, the \hiscore{} detector is being designed as a new, powerful
instrument for gamma-ray astronomy at ultra high energies. With an effective
area of 10 to $\unit[100]{km^2}$, fast read-out electronics and advanced event
reconstruction techniques it is expected to achieve a gamma-ray sensitivity
that allows a detailed study of the UHE part of the spectrum of the currently
known sources and the potential detection of many new sources in the UHE
regime. By its design it will also be a powerful instrument for the measurement
of charged cosmic rays, with a focus on the very interesting energy range
around the knee. Details of the detector can be found in \cite{martin_roma}. 

To achieve the large effective area at a reasonable effort, an inter-station
spacing of $\unit[100]{m}$ to $\unit[200]{m}$ is foreseen (to be compared with, e.g., approximately 15~m 
for the individual stations that made up the AIROBICC array). The resulting low number of data channels per event requires advanced air shower reconstruction techniques that can work with a sparse sampling of the Cherenkov light front. In this paper, the event reconstruction developed so far for this detector will be presented. All reconstruction techniques are tested using Monte Carlo simulations and the results are used to calculate the expected accuracy of the reconstruction.
 
In section \ref{simulation} the air shower and detector simulations are
introduced along with the extraction of signal parameters from the data.
Section \ref{reco} outlines the reconstruction of shower core position,
direction and energy of the primary particle and the vertical position of the
Cherenkov light maximum (shower depth). The particle separation algorithm is
presented in section \ref{separation}. The results are used in
section \ref{sensitivity} to calculate the gamma-ray flux sensitivity of the instrument.

\section{Detector simulation and measured quantities}
\label{simulation}

\subsection{Air shower simulations}
The simulation chain used for this work consists of the air shower simulation code\footnotemark{} CORSIKA \cite{heck:1998a:long}, the detector simulation \emph{sim\_score} \cite{2011AdSpR..48.1935T} and the event reconstruction \emph{reco\_score}. About 145,000 air showers with energies between $\unit[10]{TeV}$ and $\unit[5]{PeV}$ have been generated following a $dN/dE \propto E^{-1}$ spectrum. Primary particle species used include gammas, protons and helium with about 25,000 events each and nitrogen and iron with about 35,000 events each. Event core locations are scattered uniformly over the array and up to $\unit[350]{m}$ beyond the detector perimeter. Particle directions are randomised with a maximum zenith angle of 30$^\circ$. The CORSIKA IACT package \cite{2008APh....30..149B} has been used to obtain the Cherenkov light pulses at the detector level. All simulations are carried out using the US standard atmosphere \cite{us_standard_atmosphere}. The impact of varying atmospheric conditions on the results 
has not yet been studied.

\footnotetext{Using CORSIKA version 6.735 with the QGSJET module version 01c.f \citep{Kalmykov199717} for high energy hadronic interactions, the GHEISHA module \citep{fesefeldt:1985a} for lower energies, and EGS4 for the electromagnetic component.}

\subsection{Detector simulation}
The current simulations of \hiscore{} assume a regular square grid of 22 x 22 detector stations with an inter-station distance of $\unit[150]{m}$ at sea level, resulting in an instrumented area of $\unit[9.922]{km^2}$. (The effects of a different station spacing and altitude are discussed in \cite{Hampf2011_TEXAS}, and further investigations into an optimised layout, possibly using varying stations spacing, are on-going.)

Each detector station contains four 8`` PMTs equipped with Winston cone light concentrators, and has a total light sensitive area of $\unit[0.5]{m^2}$. The signals of all four channels are added up and recorded by a 1 GHz analogue-digital converter. Each station is triggered independently by signals greater than about 180 photoelectrons. Only triggered stations and stations
that have at least one triggered neighbour station are read out. The sampled signals and the corresponding time stamps are transmitted to the central data acquisition system digitally, using standard technology like ethernet or a mobile phone network.

The detector simulation \emph{sim\_score} calculates the signals seen by each station, taking into account atmospheric absorption, angle- and wavelength dependent transmission of the Winston cones, the PMT quantum efficiency, PMT
afterpulses, the signal shaping by the system's response function, and the noise in the system, which is dominated by light from night sky brightness \cite{Hampf2011_NSB}.

\subsection{Data processing}
\label{data_proc}
\begin{figure}[tb]
	\centering
	\includegraphics[angle=270, width=\columnwidth]{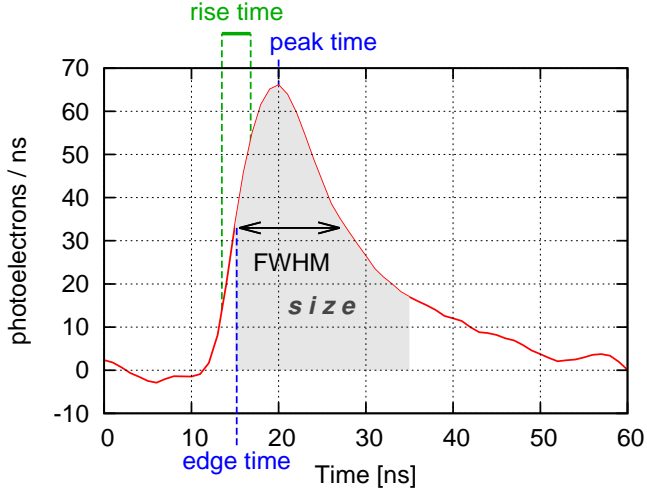}
  	\caption{Typical detector signal with illustration of the signal parameters introduced in section \ref{data_proc}. The signal is extracted from a simulated $\unit[509]{TeV}$ gamma-ray event, from a detector station about $\unit[200]{m}$ away from the shower core (summed signal of all four PMT channels). The average noise level of $\unit[90]{p.e./ns}$ has been subtracted. The x-axis shows time relative to an arbitrary reference.}
	\label{typical_signal}
\end{figure}

The signals are preprocessed for the reconstruction framework by extracting the following parameters (see also figure \ref{typical_signal}):
\begin{itemize}
 \item The time of the signal peak (\emph{peak time}), defined as the time bin with the highest entry
 \item The time where the signal reaches 50\% of the peak (\emph{edge time}), linearly interpolated between adjacent time bins
 \item The \emph{rise time}, defined as the time in which the signal rises from 20\% to 80\% of its maximum value (interpolated)
 \item The full width at half maximum (FWHM) of the signal (\emph{signal width})
 \item \emph{Signal size}, estimated by integrating the signal from 5 nanoseconds before the peak time to 15 nanoseconds after the peak time
\end{itemize}

The full sampled waveforms of the signals are stored additionally for later use in the reconstruction. 

\subsection{Quality cuts}
\label{quality_cuts}
To select a sample of good events, basic quality cuts are applied to the data (\emph{acceptance cuts}). In this study, events are used if they trigger at least three stations, have a reconstructed core position within the array and a reconstructed zenith angle of no more than $25^\circ$. This last requirement determines the effective field of view to $\unit[0.6]{sr}$. Loosening these cuts would deteriorate the reconstruction accuracy, but also improve the event statistics, and may be beneficial in some cases -- this is however not studied here. On top of the acceptance cuts, \emph{gamma-ray cuts} are used to select a sample of gamma-ray like events. These cuts are described in section \ref{separation}.

\section{Event reconstruction}
\label{reco}
The event reconstruction starts with the reconstruction of the shower core position and the direction of the primary particle as pure geometric parameters. Subsequently, the energy and the shower depth are reconstructed using lookup tables. The reconstructed values are then used for the gamma hadron separation (see section \ref{separation}). 

In the following, the resolution of each reconstructed quantity will be given as the value at which 68\% of the events are contained. All gamma-ray events from the dataset described in the previous section that survive the acceptance and the gamma-ray cuts are used to calculate the resolution. Generally, all resolutions improve with the number of triggered stations, and hence with energy of the primary particle, and will therefore be given as function of energy. Only data points above $\unit[50]{TeV}$ are presented, since the reconstruction is not yet optimised for lower energies and deteriorates quickly in that regime. The question of the energy threshold will be discussed briefly in section \ref{sec:eff_areas}.

\subsection{Shower core position}
\label{core_reco}
The most straightforward method to reconstruct the position of the shower core (the intersection of the shower axis with the detector level) is a centre of gravity calculation using signal sizes of all triggered stations, as used e.g.\ in the AIROBICC experiment \cite{1995APh.....3..321K}. This is a robust method that usually gives a good approximation of the core position even if only a few stations have triggered.

If signals from at least five stations are available, a better core position can be obtained by fitting the lateral light density function (LDF) to the recorded intensities, as suggested in \cite{2009NuPhS.190..247T}. The LDF is being parameterised as an exponential function near the shower core and a power law at larger distances, with the break between the two at $c_{LDF} \approx \unit[120]{m}$ \cite{2001APh....15...49F}:

\begin{equation}
   \mathrm{LDF}(r) = 
   \begin{cases}
      	P \, \exp(d \, r) & \mbox{for } r < c_{LDF} \\
	Q \, r^k & \mbox{for } r > c_{LDF} \\ 
   \end{cases}     
   \label{LDF}
\end{equation}
with 
\begin{align}
	r &= r(x, y) = \sqrt{x^2 + y^2}  \\
	Q &= \frac{P \, \exp(d \, c_{LDF})}{(c_{LDF})^k}  
\end{align} 

The free parameters of the fit are the absolute normalisation $P$, the inverse decay-length of the exponential function $d$, the power law index $k$ and the position of the shower core $(x, y)$. If at least six detector signals are available, $c_{LDF}$ can be a free fit parameter as well, which has been found to slightly improve the core resolution.

\begin{figure}[tb]
	\centering
	\includegraphics[angle=270, width=\columnwidth]{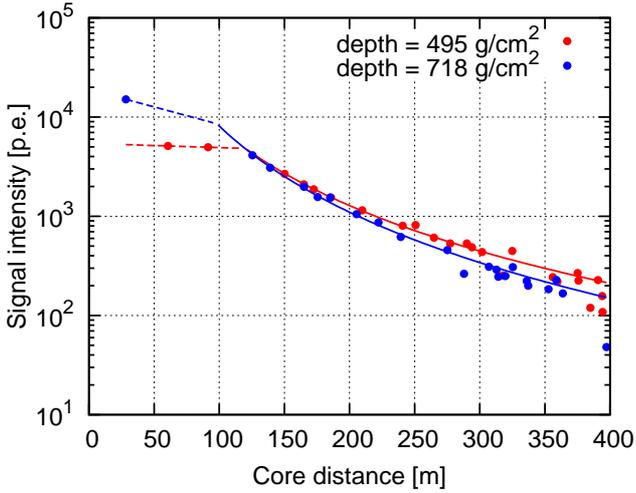}
  	\caption{The LDF of two $\unit[750]{TeV}$ gamma-ray events with different shower depths. The distributions are fitted with equation \ref{LDF}. The dashed lines denote the exponential part of the LDF, the solid lines the power law part.}
	\label{LDF_example}
\end{figure}

The data is fitted using an iterative $\chi^2$-minimisation. The data point with the largest deviation from the fit (in terms of $\upsigma$) is removed, and a new fit is conducted using the reduced data sample. This procedure is repeated until there are no more outliers further away than $1 \upsigma$ from the fit, or the number of remaining data points drops below five. The fit is considered successful if it converges numerically, the resulting $\chi^2$ value is not used for further evaluation. The same procedure is used for all following fits as well.

After the reconstruction of the shower direction (see next section), the station positions can be transformed into the shower plane (the plane perpendicular to the shower axis), and the core position fit can be repeated using the new coordinates, which yields a slight improvement especially for events with large zenith angles. Figure \ref{LDF_example} shows two examples of the LDF fitted to simulated data (two-dimensional projection). The use of the LDF for the energy and shower depth reconstruction is discussed in sections \ref{energy_reco} and \ref{depth_reco}.

\begin{figure}[tb]
	\centering
	\includegraphics[angle=270, width=\columnwidth]{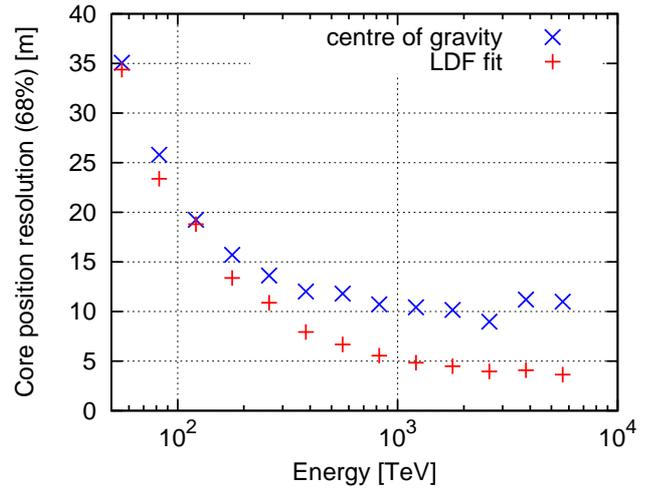}
  	\caption{Resolution (68\% containment) of the shower core position reconstruction with the two methods described in section \ref{core_reco}.}
	\label{core_res}
\end{figure}

Figure \ref{core_res} presents the resolution of the shower core position reconstruction for the centre of gravity method and the LDF fit method. The resolution is about $\unit[35]{m}$ near the threshold and improves to less than $\unit[10]{m}$ at higher energies.

While the centre of gravity method can (per definition) only reconstruct core positions within the array, the LDF fit can also be used to reconstruct core positions outside of the array. However, this method gets inaccurate quickly when the shower core moves away from the array border. At $\unit[1]{PeV}$, the resolution deteriorates from $\unit[5]{m}$ for contained events to $\unit[15]{m}$ for core positions up to $\unit[150]{m}$ away from the detector perimeter, and to about $\unit[35]{m}$
for showers between $\unit[150]{m}$ and $\unit[300]{m}$ away. 

A better method for these showers is the usage of the \emph{signal width}, which can be used up to far distances from the array as long as enough light is received in the detector to accurately measure the signal widths. This method has successfully been tested by the AIROBICC collaboration \cite{Henke_diplom} and is also used at the TUNKA detector
\cite{2010arXiv1003.0089B} to increase its effective area. This method has however not been pursued in this study, and only events that are reconstructed to be inside of the array are used for further analysis.

\subsection{Direction of origin}

\label{direction_reco}
A coarse estimate of the shower direction can be obtained by fitting a plane to the measured light arrival times\footnote{Either peak times or edge times can be used. Using the edge times results in a slightly better accuracy, as they can be measured with higher resolution.}. If signals from only three stations are available, only this estimate is used.

If at least four signals are available, the direction reconstruction can be improved significantly by taking into account the curvature of the Cherenkov light front. For this, the light arrival time model developed by \cite{2011APh....34..886S} for timing stereoscopy at Cherenkov telescopes has been adapted for an array detector. The expected time delay $t_{Det}$ at a
given detector -- with respect to the arrival time at the core position, $t_0$ -- is parameterised as function of the detector position, the height of the shower maximum $z$, and the direction of the shower axis (given by the zenith angle $\theta$ and the azimuth angle $\phi$). The detector position is given relative to the shower core in polar coordinates, using its distance $r$ and the azimuth angle $\phi_{Det}$.

\begin{equation}
t_{Det} = t_{Det}(r, \phi_{Det}, z, \theta, \phi) + t_0
\end{equation}

Using $r$ and $\phi_{Det}$ from the previous reconstruction of the shower core position, the function can be fitted to the measured arrival times, which yields the height $z$ and the shower direction.

The functional dependence is derived by integrating the light path from the point of Cherenkov light emission to the detector, using 

\begin{equation}
 \eta(h) = 1 + \eta_0 \exp(-h/h_0)
\end{equation}
for the height dependent refractive index of air ($h_0 = \unit[8]{km}$,
$\eta_0=2.76 \times 10^{-4}$). The integration yields 

\begin{equation}
  t_{det}(k, z) = \frac{1}{c} \sqrt{k}  \left( 1 + \frac{\eta_0 h_0 (1 - \exp(-z / h_0)) }{z}   \right) + t_0
  \label{eq:time_delay_long}
\end{equation}
with 

\begin{equation}
 k = k(r, \phi', z, \theta) = r^2 + \frac{z^2}{\cos^2(\theta)} - 2 r z \tan(\theta) \cos(\phi')  
 \end{equation}
with $\phi' = \phi_{Det} - \phi$ (see \cite{daniel_thesis} for details on the derivation).

\begin{figure}[tb]
\centering
\includegraphics[angle=270, width=\columnwidth]{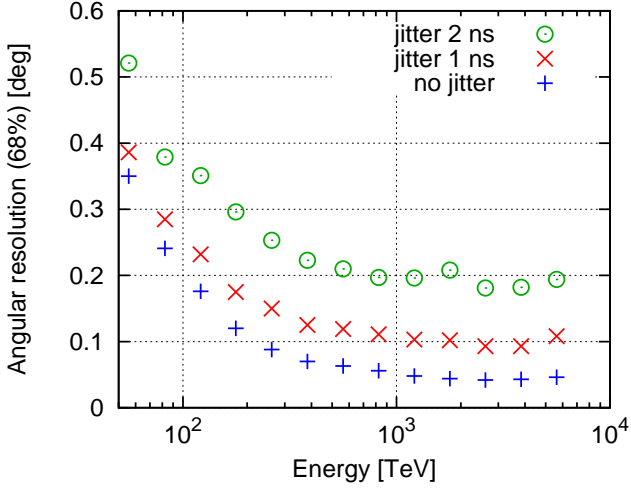}
\caption{Resolution (68\% containment) of the direction reconstruction. Three different scenarios are assumed for the time synchronisation of the signals between the stations: An ideal synchronisation, a synchronisation with a gaussian error with $\sigma = \unit[1]{ns}$ and one with $\sigma = \unit[2]{ns}$.}
\label{angres}
\end{figure}

This model is fitted to the measured arrival times (edge times) with the height $z$ and the direction described by $\theta$ and $\phi$ as free parameters. The fit results in an accurate reconstruction of the particle direction and an estimate for the height of the shower maximum. Other methods to reconstruct the shower maximum are described in section \ref{depth_reco}. 

A good angular resolution can only be achieved if the signals of the different stations can be synchronised to each other with high precision, which is a challenging experimental task given the large distances between the stations. To examine the impact of a non-ideal time synchronisation, the simulated signals have been shifted randomly in time using a Gaussian distribution with a width of $\sigma$. Figure \ref{angres} shows the angular resolution as function of energy for a perfect synchronisation and for jitters of $\sigma = \unit[1]{ns}$ and $\sigma = \unit[2]{ns}$. The angular resolution is given as the 68\% containment region of the angular distance between reconstructed and true direction of individual events.

The results show that the accuracy of the direction reconstruction is limited by the time synchronisation even if the jitter is as low as $\unit[1]{ns}$. Therefore, a time synchronisation accuracy of at least $\unit[1]{ns}$, preferably better, should be aimed at in the detector development. In the following, a time synchronisation with a jitter of $\sigma = \unit[1]{ns}$ will be assumed. With this assumption the resolution of the direction reconstruction ranges from 0.4$^\circ$ near the threshold down to 0.1$^\circ$ at higher energies.

\subsection{Energy}
\label{energy_reco}

In general, the amount of Cherenkov light generated is proportional to the energy of the primary particle. Since only a small fraction (about 0.002\%) of the total Cherenkov light arriving at the observation level is captured by the \hiscore{} detector stations (due to their large spacing and their small light collection areas), the fitted LDF is used for an estimate of the total Cherenkov light. The actual energy reconstruction is complicated by the fact that the light distribution on the ground depends strongly on the shower depth: The lower a shower maximum occurs in the atmosphere, the steeper the LDF will be,
with more light closer to the shower core and less further away (see also figure \ref{LDF_example}).

\begin{figure}
\centering
\includegraphics[angle=270, width=\columnwidth]{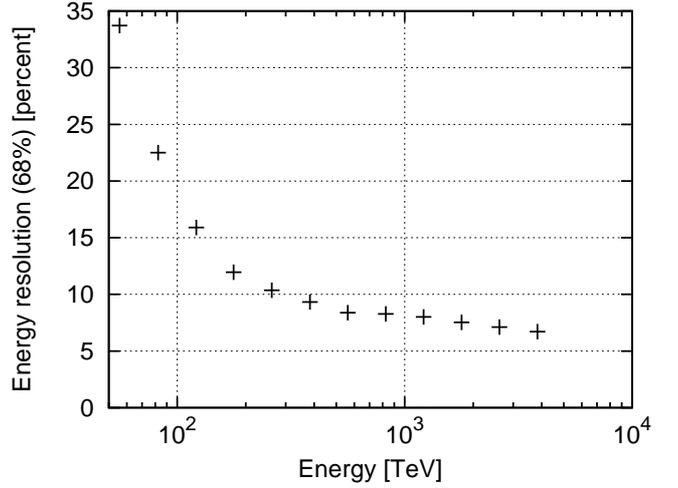}
\caption{The resolution (68\% containment) of the energy reconstruction as described in section \ref{energy_reco}}
\label{eres}
\end{figure}

The simulations carried out show empirically that the impact of the shower depth on the light distribution at detector level is minimal at a distance of around $\unit[220]{m}$ from the shower core. Therefore, the fitted light intensity at
$\unit[220]{m}$, i.e.\ LDF($\unit[220]{m}$), is used for energy reconstruction. Simulated gamma-ray events are used to generate a lookup table of $\log(LDF($\unit[220]{m}$))$ to $\log(E)$ with six bins per decade. The energy is reconstructed from this lookup table using interpolation. Figure \ref{eres} shows the resulting relative energy resolution. It improves from about 30\% near the threshold to about 10\% at higher energies. The selection bias that usually leads to an overestimation of particle energies near the threshold is negligible above $\unit[50]{TeV}$.

The energy reconstruction is optimised for gamma-ray events. Hadronic cosmic
rays produce less Cherenkov light than gamma-rays with the same energy, and
heavier hadronic primaries produce less light than lighter ones. The energy of
cosmic rays, especially heavier nuclei, is therefore always underestimated. To
derive a cosmic ray energy spectrum, an assumption about the composition must
be made.

\subsection{Shower depth}
\label{depth_reco}
The shower depth is defined as the atmospheric depth (measured from the top of
the atmosphere along the shower axis) of the maximum of Cherenkov light
emission, which (almost) coincides with the maximum of relativistic particles in the
shower. Although not of immediate interest, it is an important quantity for the
gamma hadron separation (see section \ref{separation}) and the determination of
the mass (in case of a hadronic primary). Three different (partly correlated)
methods are used here for the estimation of the shower depth:

\paragraph{Arrival time method} As shown in section \ref{direction_reco}, the
height of the light emission has an impact on the shape of the arrival time
distribution over the array. Generally, the delay of the photon arrival at
stations at a given distance from the shower core increases with shower depth.
Using the peak times of the Cherenkov light signals, the height can be inferred
from the fit to the function $t_{Det}$ (eq. \ref{eq:time_delay_long}). In
practice, the reconstruction has proven to be more accurate if the edge times
are used instead of the peak times. A lookup table (fitted value of $z$ to
shower depth) is used to correct for the shift introduced by using the edge
times. 

\begin{figure}[tb]
	\centering
	\includegraphics[angle=270, width=\columnwidth]{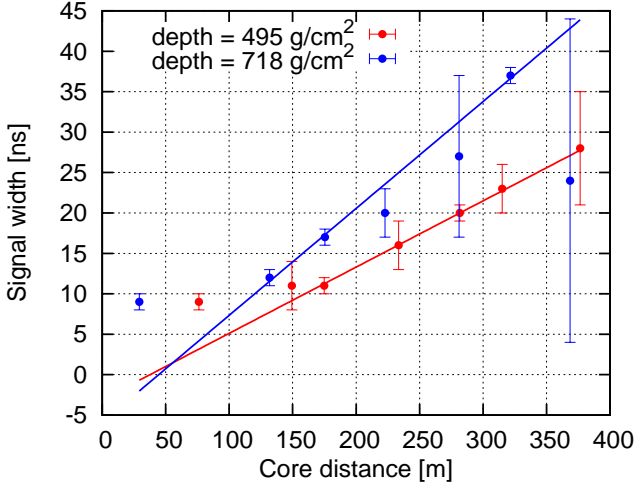}
	\caption{The signal widths of the same two gamma-ray events as in
figure \ref{LDF_example}, after summing up signals from stations within
$\unit[50]{m}$ bins. Linear fits are applied in the core distance range from
$\unit[150]{m}$ to $\unit[400]{m}$. The errorbars are estimated on basis of the
signal size after stacking.} \label{widths_example}
\end{figure}

\paragraph{LDF slope method}
As illustrated in figure \ref{LDF_example}, the Cherenkov light of showers with
a large shower depth is concentrated closer around the shower core position.
Usual parameters for the measurement of this feature are the slope of the inner
part of the LDF or the ratio of the LDF values at a small and a large core
distance \cite{1983JPhG....9.1433P, 1989JPhG...15..893D, 1998APh.....8..235L,
2001APh....15...49F, 2005ICRC....6..257B}. Since the slope of the inner part of
the LDF is not well defined in \hiscore{} due to the large station spacing, the
ratio $LDF(\unit[50]{m}) / LDF(\unit[220]{m})$ is used. Tests with other
ratios, e.g.\ $LDF(\unit[50]{m}) / LDF(\unit[150]{m})$, have yielded similar
results (the best ratio also depends slightly on the energy range). Lookup
tables are used to derive the depth from the measured ratio.

\paragraph{Signal widths method}
The width (or duration) of the Cherenkov light pulse -- defined here as full
width at half maximum, FWHM -- increases with core distance $r$. Different
parameterisations for \emph{width}$(r)$ have been suggested, e.g.\ a power law
\cite{1977ICRC....8..244K} or an exponential function
\cite{2009NuPhS.190..247T}, but for the current study a simple linear function
(as also used by \cite{0305-4616-4-9-007, Henke_diplom}) has been found to
describe the simulated data best. Close to the shower core the pulse is shorter
than the time resolution of the simulated PMT, so that this functional
dependence is only visible in detectors further than about $\unit[150]{m}$ away
from the shower core.

At a given core distance, the signal width increases with shower depth, as
observed already experimentally by \cite{0305-4616-4-9-007,
1979ICRC....9...73K} and studied in simulations by \cite{1983JPhG....9..323P}.
It can therefore be used for the determination of the shower depth, as done
e.g.\ in the TUNKA analysis \cite{ICRC2009_0492}. The sensitivity to the shower
depth increases with core distance, but the determination of the width becomes
more and more inaccurate at large core distances due to the decreasing signal
sizes, so a trade-off must be found.

To improve the measurement of the signal FWHM at large core distances, all
signals within a certain distance interval from the shower core are summed up
before determining the signal width (\emph{signal stacking}). As the number of
detector stations available in a distance interval is proportional to the
distance to the shower core, this procedure makes it possible to sample the
width up to high distances from the shower core. The distance intervals used
here are $\unit[50]{m}$ wide. Figure \ref{widths_example} shows the resulting
widths distribution of two events with different shower depths, along with
linear fits in the core distance range from $\unit[150]{m}$ to $\unit[400]{m}$. 

The depth reconstruction is done here using the fitted FWHM at $\unit[300]{m}$
in combination with lookup tables.

\paragraph{Combination of methods and performance}

\begin{figure}[tb]
\centering
\includegraphics[angle=270, width=\columnwidth]{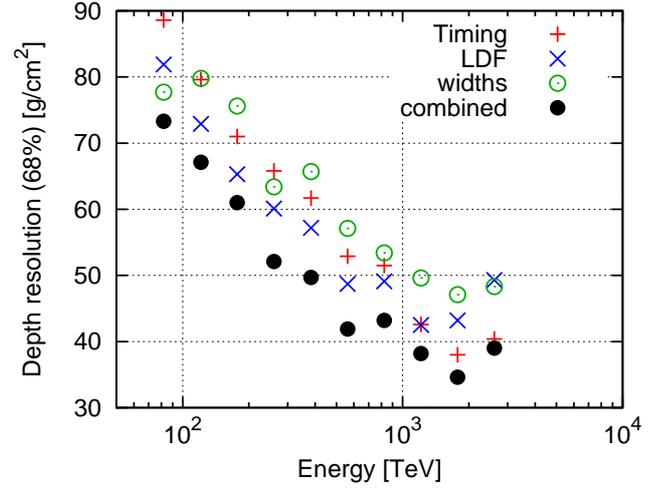}
\caption{Shower depth resolution of proton events for the three methods described in section \ref{depth_reco}, and the combination of the three methods, versus the true (simulated) energy.}
\label{depth_res}
\end{figure}

To combine the three methods, the average of all methods that return a successful depth reconstruction is used (a method may be unsuccessful if the required parameter distribution cannot be fitted due to missing or contradictory data). Figure \ref{depth_res} shows the resolution of the three individual methods and the average for a sample of proton events. Since all
methods require a good determination of parameter distributions up to large core distances, a depth estimate is difficult near the threshold where only a few stations produce a usable signal. Towards PeV energies, the resolution improves to below $\unit[40]{g/cm^2}$.

The simulations show that a particle dependent bias exists in all three depth reconstruction methods. Therefore, the depth reconstruction can only be optimised to one particle species; for the current study, it is calibrated to protons. The causes for this behaviour and the resulting potential for gamma hadron separation are discussed in section \ref{depth_bias_sep}.

\section{Gamma hadron separation}
\label{separation}

The objective of gamma hadron separation is the statistical separation of gamma-ray induced air shower events among the (more or less) constant and isotropic abundance of (charged) cosmic ray events\footnote{Since the flux of electrons decreases more rapidly with energy than the one of nuclei, electron-induced air showers are not a relevant cause of background for UHE gamma-ray astronomy. Therefore, only air showers of hadronic origin must be filtered out.}.

Since the observable differences between photonic and hadronic air showers are rather subtle, it is traditionally a difficult task to distinguish them with any air Cherenkov detector, but particularly with non-imaging detectors. Nevertheless the separation is possible to some extent, as will be shown.

It should be noted that small gamma-ray emission regions, especially point-like sources, can be detected without any gamma hadron separation, as they emerge from the isotropic flux of background as a localised excess. Nevertheless, a good gamma hadron separation improves the significance of a detection, or in other words, enables the detection of weaker gamma-ray sources during the same observation time. 

In the following, different methods for particle separation will be discussed qualitatively. The combination of separation parameters and the resulting quantitative performance will be discussed section \ref{gh_performance}.

\subsection{Separation using depth reconstruction bias}
\label{depth_bias_sep}
The simulations show that all three described depth reconstruction methods exhibit a particle dependent bias. With the LDF and the widths method, the depth of showers induced by heavier hadrons (e.g.\ iron) is systematically overestimated by 60 to $\unit[100]{g/cm^2}$, while the depth of photonic showers is underestimated by 20 to $\unit[40]{g/cm^2}$. Since the lookup tables are done with a proton sample, no bias exists for protons. If the arrival time method is used, the offset between gamma-rays and iron nuclei, the most extreme cases among the simulated species, amounts to only 20 to $\unit[50]{g/cm^2}$.

The reason for these offsets are differences in the shower development below the shower maximum that depend on the type of primary particle. Air showers induced by hadrons, especially heavier nuclei, contain large numbers of secondary hadrons. These particles do not loose energy as rapidly as particles from the electromagnetic cascade and therefore penetrate deeper into the atmosphere. Even at low altitudes, they constantly refuel (e.g.\ by pion decay) the electromagnetic component of the shower, which results in more Cherenkov light closer to the ground. Since the LDF and widths methods are based upon the full sampled Cherenkov light pulses, they are sensitive to all stages of the shower development. The Cherenkov light from low altitudes shifts the reconstructed shower maximum to there, i.e.\ to larger depths. The arrival time method, on the other hand, is only sensitive to the Cherenkov light emitted at a specific point of the shower development (e.g.\
the shower maximum, if the peak times are used) and is therefore not influenced by the differences between the shower types.

There are several ways to exploit these facts for particle separation. Here, the difference between the depth reconstructed with the widths method, $X_{width}$, and the arrival time method, $X_{timing}$, is used. The centre of the distribution of  $X_{width}-X_{timing}$ is around zero for protons (since the lookup tables for both methods are done with a proton sample), below zero for gammas and above zero for heavier nuclei. Although the distributions overlap significantly, a particle separation can be achieved to some extent.

It should be noted that the simulations indicate that the offsets between the reconstructed depth values are most pronounced for a detector at sea level and almost vanish at an altitude of $\unit[2000]{m}$, probably because the later stages of the shower, which are sensitive to the hadron content, do not even develop above this altitude. For the same reason, events with higher energies (starting at low PeV energies) show smaller offsets, as their shower depth is larger on average and the later stages of the shower can no longer develop above the detector level. This may explain why such a bias has not yet been
noted by other non-imaging air Cherenkov arrays (e.g.\ AIROBICC \cite{1998APh.....8..235L}, BLANCA \cite{2001APh....15...49F} or TUNKA \cite{ICRC2009_0492}): Either their altitude was too high, or the focus was on too large energies, or both.

\subsection{Separation using shower depth versus energy}
\label{sep:depth_vs_energy}
The depth of an air shower is largely determined by the altitude of the first interaction of the primary particle in the atmosphere. The altitude of the first interaction, in turn, depends on the cross section of the reaction between the primary particle and air molecules. The shower depth increases logarithmically with energy and is larger for photonic than for hadronic events. Additionally, heavier nuclei result in a smaller shower depth than lighter nuclei at the same energy. 
Therefore, the shower depth (in combination with the previously reconstructed particle energy) can be used as an indication of the particle type. This method is widely used in the field of air shower detectors, usually to derive the mean mass of charged cosmic ray particles (see e.g.\ \cite{2009PrPNP..63..293B}). 


The particle identification is complicated by the fact that the depth of hadrons, especially heavier hadrons, is usually overestimated by the described depth reconstruction methods. To keep the bias on the reconstructed shower depth as small as possible, only the arrival time method is used here. Additionally, the energy of hadronic particles is underestimated. Both offsets decrease the observable differences between the particles. The remaining subtle differences between nucleonic and gamma-induced air showers are washed out to some extent by the intrinsic fluctuations of the air shower development. Nevertheless, some separation of gammas and hadrons can be achieved. 

\begin{figure}[tb]
\centering
\includegraphics[angle=270, width=\columnwidth]{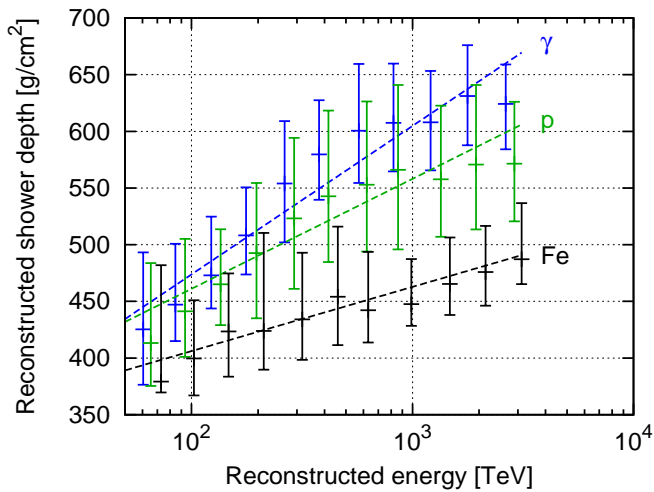}
\caption{Median reconstructed depth (arrival time method) versus reconstructed energy for gammas, protons and iron nuclei. The errorbars denote statistical variation (68\% containment), i.e.\ include both the shower fluctuations and the uncertainty of the reconstruction. Values for protons are shown at 10\%, for iron nuclei at 20\% higher energies for clarity.}
\label{depth_vs_energy}
\end{figure}

Figure \ref{depth_vs_energy} shows the median reconstructed depth values for different particles as function of reconstructed energy. The median reconstructed depth for photonic events is parameterised as function of energy. The deviation of a measured value from this expected depth is used as particle separation parameter. While the overlap between the distributions for gammas and protons is considerable, heavier nuclei can be identified and discarded rather efficiently.

\subsection{Separation using signal rise time}
\label{rise_time}

As noted in section \ref{depth_bias_sep}, air showers of hadronic origin contain more secondary hadrons which refuel the electromagnetic cascade down to low altitudes. Since these secondary hadrons move faster than the speed of light in air, the Cherenkov light produced by the corresponding electromagnetic cascade appears in the detector \textit{before} the bulk of Cherenkov light generated from particles near the shower maximum. Therefore, hadronic events can be identified by some "early light" before the peak of the signal.

\begin{figure}[tb]
\centering
\includegraphics[angle=270, width=\columnwidth]{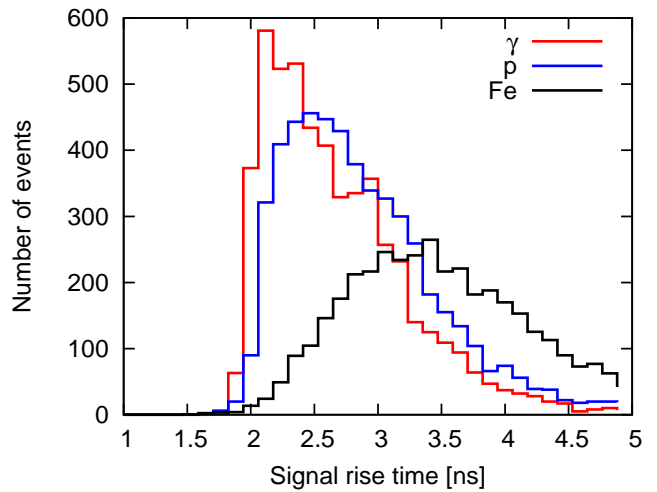}
\caption{Distribution of the signal rise time of the central detector station for gammas, protons and iron nuclei with (true) energies between $\unit[100]{TeV}$ and $\unit[1]{PeV}$.}
\label{rise_time_histo}
\end{figure}

Here, this feature is detected by using the signal rise time (see section \ref{data_proc}). For hadrons (especially heavier ones), a longer rise time is expected. The effect is most pronounced near the shower core, since at larger core distances the light from low altitudes must cross a significantly longer distance through air and no longer appears before the main peak. Therefore, only the signal from the ''central station'' (the station closest to the shower core) is used. Figure \ref{rise_time_histo} shows the distribution of this parameter for various particles and the main energy range of \hiscore{}. While again the distributions overlap considerably, a separation is evident.

The rise time has been suggested previously as particle separation parameter for Cherenkov telescope data (see e.g.\ \cite{2002APh....17..497R}), but due to its close correlation with the image shape recorded in the telescopes it was considered not useful \cite{1997APh.....6..343A}. It seems however that for non-imaging detectors the rise time can be a useful parameter for gamma hadron separation.

\subsection{Separation using Cherenkov light spectrum}
\label{spectral_sep}
The Cherenkov light emitted by hadronic air showers at low altitudes can not only be identified by its time signature, but also by the observable light spectrum. Since UV light (around $\unit[250]{nm}$) is strongly absorbed by air, its presence at the detector level is an indication of light emitted at low altitudes, and thus of hadronic events.

The use of this method has been suggested previously by \cite{1983IzKry..66..234S}, and applied to simulations of imaging Cherenkov telescopes by \cite{1991NIMPA.302..522A} and \cite{2002APh....17..497R}. Although a signature of the particle type could be found in the amount of measured UV light, these studies concluded that other, easier to measure, parameters contain the same or even better information.

However, these studies concentrated on imaging Cherenkov telescopes, and it seems plausible to assume that the spectral information may be more useful in a non-imaging detector (in a similar way as the rise time, see section \ref{rise_time}).
Therefore, the effect of a spectral measurement for \hiscore{} was tested by adding an additional UV sensitive channel to the standard detector stations in the simulation. The amount of detected UV light is normalised to the light detected in the standard detection channel, yielding the \emph{UV light ratio}. 

The simulations show that almost no particle separation can be achieved by this method, as two contrary effects cancel out each other: On the one hand, the UV light ratio contains a signature of the shower depth. Since air showers of \emph{photonic} origin penetrate on average deeper into the atmosphere than hadronic showers (see e.g.\ figure \ref{depth_vs_energy}), they contain more particles close to the detector level, which increase the UV light ratio. On the other hand, \emph{hadronic} events exhibit more deeply penetrating secondary hadrons which produce some early light close to the detector (see explanation in sections \ref{depth_bias_sep} and \ref{rise_time}) and by that a surplus of UV light.

A correction for the (reconstructed) shower depth can be used to isolate the latter effect, but introduces additional uncertainties. The remaining differences between different particle species are much smaller than in the previously discussed methods. Therefore, the additional effort of a spectral measurement does not seem to be justified.

\subsection{Combination of methods and performance}
\label{gh_performance}

To rate the performance of the gamma hadron cuts, the quality factor $QF$ is used:

\begin{equation}
	QF = \frac{\epsilon_\gamma}{\sqrt{\epsilon_{bg}}}
\end{equation}
where $\epsilon_\gamma$ and $\epsilon_{bg}$ are the survival probabilities for gamma-ray and cosmic ray (background) events, respectively. In the (common) case of background-dominated observations, the quality factor has a linear impact on the instrument sensitivity.

The survival probabilities for each particle species can be determined by applying the gamma hadron cuts to the simulated events. However, since all described particle separation methods discriminate heavier hadrons more efficiently than lighter ones, the correct calculation of $\epsilon_{bg}$ must assume a realistic composition of the background events. For this, the polygonato model by \cite{hoerandel:2003a} is used, which yields parametrisations of the energy spectra of all elements up to $Z=92$. The survival probability for events of each particle species are calculated using the particle separation power of the most similar simulated element (iron, nitrogen, helium, hydrogen). 

Additionally, the non-ideal energy reconstruction must be taken into account: Since the energy of cosmic rays is underestimated on average (see also section \ref{energy_reco}), the background at a given reconstructed energy comprises of cosmic ray events with a higher true energy. In order to compare $\epsilon_\gamma$ and $\epsilon_{bg}$, both are always given for reconstructed energies.

\begin{figure}[tb]
\centering
\includegraphics[angle=270, width=\columnwidth]{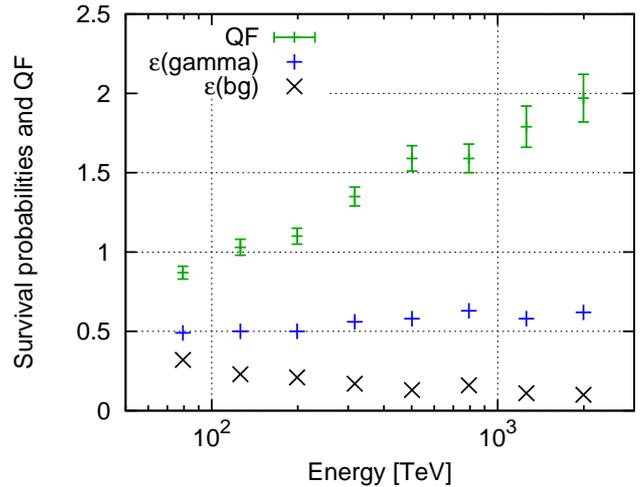}
\caption{Gamma-ray and cosmic ray (background) survival probabilities, and quality factor, after the application of the described combination of gamma hadron cuts, versus the reconstructed energy. 
The uncertainties of the survival probabilities are about the size of the markers or smaller.}
\label{QF_combi1}
\end{figure}

Several strategies can be used to combine the three particle separation methods introduced in sections \ref{depth_bias_sep} to \ref{rise_time} (the spectral method discussed in section \ref{spectral_sep} is not used). Here, the cuts on each parameter are applied successively to each event. The (energy dependent) cut values for each parameter are adjusted to leave about $80\%$ of the gamma-ray events at each stage. The resulting total gamma-ray survival probability $\epsilon_\gamma$ ranges from $50 \%$ to $60 \%$, while $\epsilon_{bg}$ improves from $30\%$ at lower energies to less than $10\%$ at higher energies (see figure \ref{QF_combi1}). The corresponding quality factor improves with energy from 0.9 near the energy threshold to about 2.0 at PeV energies. A more sophisticated way to combine the three cuts, e.g.\ a multivariate analysis, may improve the performance slightly (see e.g.\ \cite{2009APh....31..383O} for a corresponding work for imaging Cherenkov telescopes).

It should be noted that, depending on flux level, observation time and other factors, there may be an energy above which the sensitivity is limited by the number of signal events rather than by the background (background-free regime). Obviously, the gamma hadron separation should be adjusted at these energies in order to retain as many gamma-ray events as possible.

Near the energy threshold, up to a few hundred TeV, almost no gamma hadron separation can be achieved with the described strategy. The reason for this is the poor accuracy of the depth reconstruction in this regime, which deteriorates the separation of the methods described in sections \ref{depth_bias_sep} and \ref{sep:depth_vs_energy}. However, using the rise time method (section \ref{rise_time}) alone, a quality factor of 1.2 to 1.3 can be achieved from $\unit[50]{TeV}$ on. At higher energies, the rise time alone is less effective than the described combination of parameters. In practice, the most appropriate gamma hadron separation scheme should be selected after the reconstruction of the energy.

\section{Detector sensitivity}
\label{sensitivity}

\subsection{Effective areas}
\label{sec:eff_areas}
\begin{figure}[tb]
\centering
\includegraphics[angle=270, width=\columnwidth]{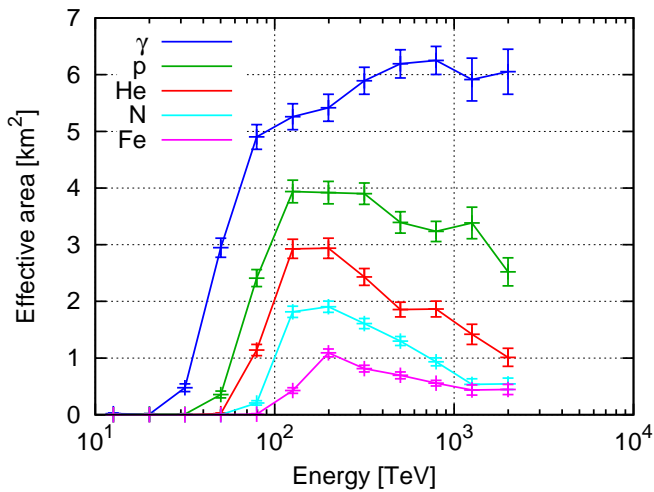}
\caption{Effective areas for all five simulated particle species after gamma hadron separation. Errorbars denote statistical uncertainties estimated from the expected Poissonian variance in each bin.}
\label{eff_areas}
\end{figure}

The effective areas of the detector are given by the instrumented area $A$ ($\unit[10]{km^2}$ for these simulations) and the average probabilities for events to survive both the acceptance ($\epsilon_{acc}$) and the gamma hadron cuts ($\epsilon_\gamma$ or $\epsilon_{bg}$, depending on the type of particle):

\begin{equation}
  A_{eff,\gamma / bg} = A \; \epsilon_{acc}  \; \epsilon_{\gamma / bg}
\end{equation}
Like the survival probabilities $\epsilon_{acc}$, $\epsilon_{\gamma}$ and $\epsilon_{bg}$, they are a function of energy and particle. Figure \ref{eff_areas} shows the effective areas as determined from the simulations, for all five simulated particle species. As expected, the effective areas for heavier nuclei are reduced efficiently by the gamma hadron separation.

From the effective areas it can be seen that gamma-rays can be detected from energies of about $\unit[30]{TeV}$. At about $\unit[50]{TeV}$, the effective area for gamma-ray detection reaches $\unit[3]{km^2}$, roughly 50\% of its maximum value.

For the study of charged cosmic rays, the analysis will be performed without the gamma hadron separation. In this case, the effective areas are equal to the instrumented area ($\unit[10]{km^2}$ in the simulated case) above a particle dependent energy (about $\unit[120]{TeV}$ for protons, about $\unit[300]{TeV}$ for iron nuclei).

\subsection{Point source sensitivity}
\label{sensitivity_detail}

The point source sensitivity of the detector is one of the key performance figures and can be compared with the flux of gamma-ray sources. It is estimated here as the minimal flux of gamma-rays that is needed to detect a point-like gamma-ray source with $5 \upsigma$ above the uniform background of cosmic ray events and at least 50 gamma-ray events (\emph{detection criteria}). 

The rate of background events is calculated using the flux parameterisations from \cite{hoerandel:2003a} together with the effective areas shown in figure \ref{eff_areas}. The systematic underestimation of the cosmic ray energy is taken into account and reduces the flux seen at a given reconstructed energy by about a factor of two.

To calculate the number of gamma-ray events required for a detection, the rate of background events \textit{in the source region} must be known. For the estimation of the sensitivity, a circular source region with a radius equal to the angular resolution shown in figure \ref{angres} is assumed\footnote{Note, that the choice of the optimum size of the source region depends on the source strength as well as its energy spectrum which are a priori not known.}. For a conservative estimate, an inter-station time synchronisation of $\unit[1]{ns}$ is assumed. For a more optimistic estimate, an ideal time synchronisation is assumed. Since the sensitivity is proportional to the square root of background events, it depends linearly on the angular resolution.

\begin{figure}[tb]
\centering
\includegraphics[angle=270, width=\columnwidth]{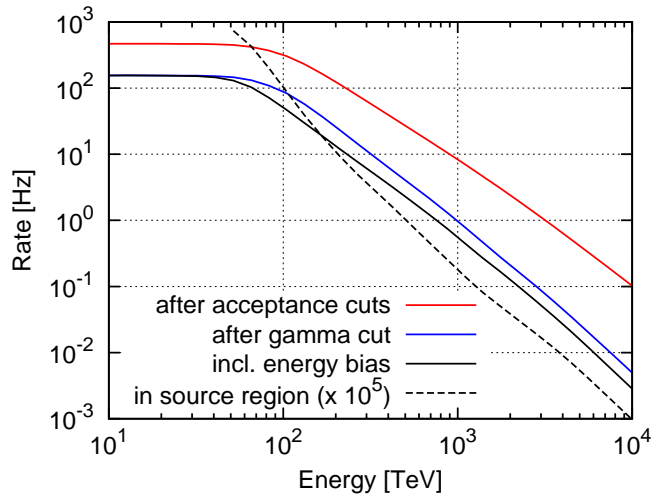}
\caption{The integral cosmic ray event rates seen by \hiscore{} before and after gamma hadron separation. The third line demonstrates the effect of the underestimated reconstructed energy for hadronic events. Lines 1 to 3 show the cosmic ray rates for the whole field of view. The fourth line (drawn at $10^5$ times its true value) gives the number of cosmic ray events (after gamma cuts and including energy bias) within the source region, which is defined by the angular resolution shown in figure \ref{angres} (using no jitter).}
\label{CR_rates}
\end{figure}

Figure \ref{CR_rates} shows the calculated rates of cosmic ray events for the whole field of view and within the source region. The values for the effective areas and the angular resolution are inter- and extrapolated where needed.


To determine the background level that needs to be subtracted from the number of events in the source region, several methods can be used \cite{2007A&A...466.1219B}. Usually, the background is measured in a sky region within the same field of view, but sufficiently far away from the source to avoid a contribution from the source itself. This background region can be larger than the source region to reduce the statistical uncertainty on the background estimate. The ratio of the solid angles of the source and the background regions is expressed by the $\alpha$-factor:

\begin{equation}
  \alpha = \frac{\Omega_{source}}{\Omega_{background}}
\end{equation}

For the conservative estimate, $\alpha = 1$ is assumed. For the optimistic scenario, $\alpha \ll 1$ is assumed, which means that the background region is chosen much larger than the source region. Due to the large field of view of \hiscore{}, and the rather flat instrument acceptance (see \cite{daniel_thesis} for acceptance plots), a low $\alpha$-factor is anticipated.

Using the background levels shown in figure \ref{CR_rates}, the number of gamma-rays with energy $E > E_0$ needed to fulfil the detection criteria is calculated for a range of values of $E_0$ between $\unit[50]{TeV}$ and $\unit[10]{PeV}$. The gamma-ray flux is modelled using a spectrum proportional to $E^{-2.6}$ without a cutoff, and the flux constant is adjusted to produce the required number of gamma-rays. 

\begin{figure}[tb]
\centering
\includegraphics[angle=270, width=\columnwidth]{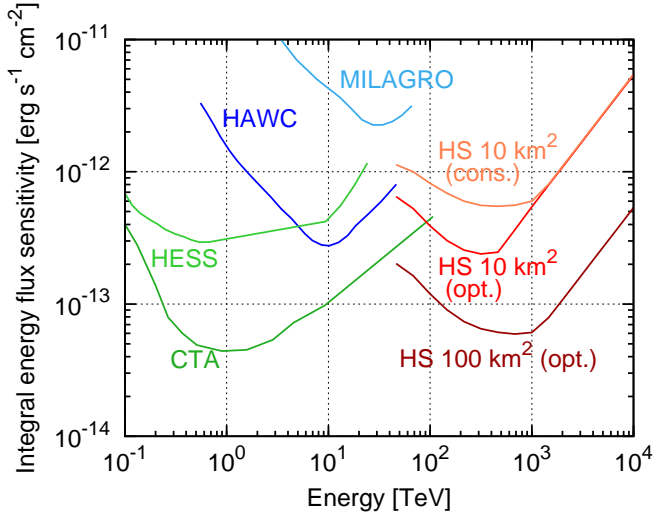}
\caption{Point source sensitivities for the simulated \hiscore{} detector ("HS"), for conservative and optimistic assumptions (see section \ref{sensitivity_detail} for details). Also shown is the expected sensitivity for a $\unit[100]{km^2}$ \hiscore{} array. For comparison, published sensitivities of selected other gamma-ray observatories are shown (CTA \cite{2011ExA....32..193A}, H.E.S.S. \cite{2008AIPC.1085..874B}, Milagro and HAWC \cite{hawc_website}). The observation time is assumed to be 50 hours for pointed instruments (H.E.S.S.\ and CTA) and five years of continuos operation for all other instruments.
}
\label{sensi_plot}
\end{figure}

Figure \ref{sensi_plot} shows the calculated sensitivity for the \hiscore{} detector in comparison with other planned or existing gamma-ray observatories. For \hiscore{}, three scenarios are plotted: A $\unit[10]{km^2}$ array with conservative assumptions ($\unit[1]{ns}$ time resolution, $\alpha=1$), a $\unit[10]{km^2}$ array with optimistic assumptions (ideal time
synchronisation, $\alpha \ll 1$), and a $\unit[100]{km^2}$ array with optimistic assumptions. Due to the wide field of view of $\unit[0.6]{sr}$, about 25\% of the sky will receive an exposure of more than 200 hours per year \cite{Hampf2011_TEXAS}. Therefore, the sensitivity has been calculated for an observation time of 1000 hours, equivalent to five years of continuous operation.

It can be seen that \hiscore{} can contribute sensitive observations in a so far poorly covered energy (or wavelength) band of the electromagnetic spectrum. A more detailed discussion of the scientific potential of an instrument with this sensitivity, also in the light of existing measurements and models, can be found in \cite{martin_roma}.

\section{Conclusions}
An event reconstruction algorithm for a non-imaging air Cherenkov array with sparse sampling has been presented, and its performance has been evaluated using a dedicated simulation framework. The choice of station spacing and detector characteristics have been guided by the \hiscore{} array, but the methods should be applicable as well to other air shower detectors of similar kind. The algorithm incorporates techniques well established in the field, but also new approaches like the arrival time model for the direction and shower depth reconstruction or the signal stacking for the measurement of the signal width. The gamma hadron separation algorithm has been newly developed, partly using ideas that have been raised in other studies. The described algorithm has been used to demonstrate that even with a sparse array such as the simulated one (station spacing $\unit[150]{m}$) an accurate event reconstruction is possible down to about $\unit[50]{TeV}$ (see table \ref{performance}). 

\begin{table}
\begin{tabular}{lccc}
  &  at $\unit[100]{TeV}$ &  at $\unit[1]{PeV}$ \\ 
 Core position res. [m] & 20 & 5  \\ 
 Direction res. [deg] & 0.25 & 0.1   \\ 
 Energy res. [percent] & 20 & 10   \\ 
 Depth res. [$\unit{g/cm^2}$] & 70 & 40   \\ 
 Sensitivity $\unit[10]{km^2}$ [$\unit{erg \, s^{-1}\, cm^{-2}}$] & $4\times 10^{-13}$ &  $5\times 10^{-13}$ \\ 
 Sensitivity $\unit[100]{km^2}$ [$\unit{erg \, s^{-1}\, cm^{-2}}$] & $1\times 10^{-13}$ & $6\times 10^{-14}$  \\ 
\end{tabular} 
\caption{Performance figures of the described reconstruction algorithm, as derived with the \hiscore{} simulations. Core, direction and energy resolutions are given for gamma-ray events, depth resolution for proton events. The \hiscore{} sensitivities are given for the optimistic scenario.}
\label{performance}
\end{table}

The results show that with the assumed configuration the \hiscore{} detector can be a powerful observatory for gamma-ray astronomy. Its sensitivity is sufficient to study the continuation of currently known gamma-ray source spectra to ultra high energies, and offers a great discovery potential for Galactic pevatrons and other sources of ultra high energy gamma radiation.

\section*{Acknowledgements}
The authors like to thank Victor Stamatescu for many fruitful discussions about the arrival time model presented in section \ref{direction_reco}.  Daniel Hampf acknowledges the financial support by the German Federal Ministry of Education and Research (BMBF contract number  05A08GU1).





\bibliographystyle{Science}


\bibliography{reco_paper}

\end{document}